\documentstyle[prb,aps,preprint]{revtex}

\title{Exchange Coupling Between Iron Layers Separated by BCC Copper}
\author{A. T. Costa Jr., J. d'Albuquerque e Castro, R. B. Muniz \\
Instituto de F\'\i sica, Universidade Federal Fluminense, Niter\'oi,
24210-340, Brazil \\ M. S. Ferreira \\ Department of Mathematics,
Imperial College, London, SW7 2BZ, U.K.\\ J. Mathon \\ Department of
Mathematics, City University, London, EC1V 0HB, U.K. }

\date{\today}

\begin{document}

\draft

\maketitle

\begin{abstract}
The exchange coupling between Fe layers separated by BCC Cu is
calculated for Fe/Cu/Fe (001) trilayers. It is shown that the
coupling is basically regulated by three extrema of the bulk BCC Cu
Fermi surface. The contributions from those extrema are all of the
same order of magnitude, but that associated with the ``belly'' at
the $\Gamma$-point dominates. The calculated temperature dependence
of the coupling varies considerably with spacer layer thickness.
Individually, the amplitudes of these extrema contribution decrease
with temperature, each according to a different rate. Such an effect
may cause an actual increase of coupling with temperature for some Cu
thicknesses.
\end{abstract}

Although the common crystal phase of bulk Cu is FCC, it is possible
to grow thin films of BCC Cu on Fe (001). The BCC stacking proceeds
for up to 12 or 20 atomic planes approximately, but for larger
thicknesses significant lattice modifications occur, leading to a
structural transformation.\cite{HeinrichR,Johnson}

The exchange coupling between Fe layers separated by BCC Cu has been
measured by groups at Simon Fraser University (SFU) and
Philips.\cite{Johnson,CH} Both have found that the coupling
oscillates with decreasing amplitude as a function of the Cu
thickness, but their results disagree in several important
aspects.\cite{HeinrichR} The Philips group data show well-defined
short-period oscillations\cite{Johnson} whereas the SFU group
originally observed a long-period oscillatory coupling.\cite{CH}
Later, the SFU group found some indication of a short-wavelength
oscillation in samples with smoother interfaces.\cite{HeinrichR} The
exchange coupling in multilayers can be strongly affected by sample
interface quality.\cite {Unguris} It is widely accepted that
interface roughness tends to suppress short-wavelength oscillations
and reduce the coupling amplitude. Therefore, as pointed out in Ref.
\onlinecite{CH}, it is rather puzzling that the values obtained at
Philips are substantially smaller than those of the SFU group.

Motivated by these apparently conflicting experimental results, we
have undertaken a theoretical analysis of the exchange coupling
between Fe layers across BCC Cu in Fe/Cu/Fe (001) trilayers. The
coupling $J$ is calculated for several temperatures $T$ and spacer
layer thicknesses $N$, using an extension of the formulation
developed in Ref.\onlinecite{jose}. For suficiently large $N$, we
divide $J(N)$ into oscillatory components coming from extrema which
are related to the spacer Fermi surface (FS). These oscillatory
contributions to the coupling are calculated separately. Our results
show that for perfectly smooth interfaces $J(N)$ is dominated by
short-period oscillations. We find that the temperature dependence of
the coupling changes significantly with spacer layer thickness.  The
amplitude of each oscillatory component decreases with temperature,
but they do so at different rates. We show that this may cause a
surprising effect which is the increase of the coupling with
temperature for some Cu thicknesses.

The interlayer exchange coupling, defined as the total energy
difference per surface atom between the antiferromagnetic and
ferromagnetic configurations of the trilayer, is given by
\cite{castro95,VersL}:

\begin{equation}
J =  \sum_{\vec{k}_{\parallel}} \int d\omega f(\omega)
F(\vec{k}_{\parallel},
\omega, N)
\label{j}
\end{equation}
where
\begin{equation}
F = \frac{1}{\pi}Im\, tr\, \ln [1 + S^{\uparrow}
(G_{mm}^{c\uparrow} -G_{mm}^{c\downarrow})
S^{\downarrow} (G_{m+1,m+1}^{c\uparrow} -G_{m+1,m+1}^{c\downarrow})],
\end{equation}
$S^\uparrow = t^\dagger
(1-G_{mm}^{c\uparrow}tG^{c\uparrow}_{m+1,m+1}t^\dagger)^{-1}$ and
$S^\downarrow = t(1-G_{m+1,m+1}^{c\downarrow}t^\dagger
G^{c\downarrow}_{mm}t)^{-1}$. In the equations above
$\vec{k}_{\parallel}$ are the wave vectors parallel to the layers,
$f(\omega)$ is the usual Fermi-Dirac distribution function, and $m$
is a plane index. As in Ref.\onlinecite{VersL} we consider an
imaginary cleavage plane across the spacer between planes $m$ and
$m+1$, separating the trilayer into two semi-infinite systems.
$G_{mm}^{c\sigma}$ and $G_{m+1,m+1}^{c\sigma}$ are matrices in
orbital indices representing the surface one-electron Green functions
of the left and right cleaved systems, respectively. The trace is
taken over orbital indices and $t$ denotes the spacer hopping matrix.

This formula for the coupling is an extension of the result
previously obtained in Ref. \onlinecite{jose} and, for the one band
model, reduces to the torque formula of Edwards et al.
\cite{drm} In deriving it we have assumed that the electrons are
noninteracting in the spacer and experience exchange-split
one-electron potentials in the ferromagnetic layers. Most of the
experimental results are for the bilinear exchange coupling term
$J_{1}$ which, for perfectly smooth Fe/Cu (001) interfaces, is
virtually equal to $J/2$.\cite{jose}

We have calculated the required Green functions within the
tight-binding model with $s,p,d$ orbitals and hopping to second
nearest neighbours.  The tight-binding parameters for all BCC Cu
planes were determined from a first-principles LMTO-tight-binding
electronic structure calculation of paramagnetic bulk BCC Cu. The
parameters for ferromagnetic Fe were obtained from paramagnetic bulk
Fe\cite{Victora}, by self-consistently adjusting the on-site
energies, assuming charge neutrality. The effective intra-atomic
electron-electron interactions were taken to be $U_{sp}^{Fe}=0$ and
$U_{d-d}^{Fe}=1 eV$. \cite{Cooke,himpsel} We neglect atomic potential
differences due to the magnetic configuration change, thus making the
approximation known as the ``force theorem''. The
$\vec{k}_{\parallel}$ sum in Eq.(1) is performed numerically and the
energy integral is evaluated in the complex plane by summing over
Matsubara frequencies at finite $T$.

The calculated results at $T$=300K for the bilinear exchange coupling
$J_1$ as a function of Cu thickness are presented in Fig.\ref{fig1} (full
circles).  Our results clearly show a short-period oscillatory
exchange coupling, in excellent agreement with the Philips group data
as far as the period of oscillation is concerned. 

For sufficiently large spacer thickness it is possible to express the
coupling as a sum of oscillatory components whose periods are
determined by extrema that are related to the spacer FS.
\cite{Bruno,PRL1,edwards94,mauro,stiles} It is essential to use a
non-perturbative treatment, as in the quantum well approach, to
analyze the relative importance of these contributions.
\cite{mauro,PRL2}  They depend upon the degree of
confinement experienced by the carriers of both spin orientations in
the corresponding extremum states, in the ferromagnetic and
anti-ferromagnetic configurations of the magnetic
layers.\cite{VersL,mauro,PRL2,MME1} The widespread practice of
considering only the periods predicted by RKKY theory, and treating
the amplitudes and phases of these contributions as adjustable
parameters may be inadequate and rather misleading. The fitting
usually involves several parameters and, in some cases, is not
unique.  Besides, when the spacer FS has to be regarded as consisting of
more than one sheet, periods not predicted by RKKY theory may
exist.\cite{mauro} Moreover, at finite temperatures, the decrease of
the oscillation amplitudes as $N$ increases is different for each
extremum and may deviate strongly from the $(1/N^2)$ asymptotic
regime\cite{VersL,PRL2,TD} usually assumed in that sort of fitting.

To identify the periods of oscillations of $J(N)$ it is useful to
look at the spacer FS.  In BCC Cu only one energy band
$E(\vec{k}_{||},k^{\perp})$ crosses the Fermi energy $E_F$. Its
calculated FS, shown in Fig.\ref{fig2}(a), is basically a sphere with
twelve ``necks'' developing at each face centre of the bulk BCC first
Brillouin zone. In the (001) direction of growth, three sets of
$\vec{k}^0_{||}$ associated with the FS extrema contribute to the
coupling. The first set consists of a single wave vector
$\vec{k}^0_{||}=(0,0)$ ($\Gamma$-point) related to the FS ``belly''.
The other two are associated with the ``necks''.  Set 2 consists of
four vectors $\vec{k}^0_{||}$: $(0,\pm 0.327)$ and $(\pm 0.327,0)$,
and set 3 of the $M$-points located at $(\pm 0.5,\pm 0.5)$. Here all
wave vectors are given in units of $2\pi/a$ where $a$ is the lattice
constant.  Due to the layered stucture of the system, it is useful to
work with the layer adapted bulk Brillouin zone instead of the usual
BZ.  The former is defined as a prism whose base is the
two-dimensional first BZ and whose height is $2\pi/d$, where $d$ is
the interplane distance perpendicular to the layers. The relevant
cross sections of the spacer FS, together with the corresponding
extremal wave vectors $\vec{k}^{\perp}(\vec{k}_{||}^0,E_{F})$, are
shown in Fig.\ref{fig2}(b). We must distinguish sets 1 and 2 from set
3 because, for the latter, the FS can be regarded as consisting of
more than one sheet. This is because more than one extrema occurs in
the first prismatic bulk BZ for each wave vector $\vec{k}^0_{||}$ of
set 3.

Considering that the integrand $F$ in Eq.\ref{j} is an oscillatory
function of $N$ we can expand it in a Fourier series. However, it is
necessary to generalize the expansion to a multiple Fourier series
\cite{castro95,mauro}, when the equation
$E(\vec{k}_{||},\vec{k}^{\perp})=\omega$ has more than one pair of solutions
$\pm k_\xi^\perp(\vec{k}_{||},\omega)$. In this case, the general
expansion of $F$ is

\begin{equation}
F(\vec{k}_{\parallel},\omega,N)=\sum_{n_1,\dots,n_\xi}c_{n_1,\dots,n_\xi}
(\vec{k}_{\parallel},\omega)e^{i\sum_{\xi} n_\xi k_\xi^\perp
(\vec{k}_{\parallel},\omega)Nd}
\label{j2}
\end{equation}

For $N \gg 1$ the exponential in Eq.\ref{j2} oscillates rapidly as a
function of $\vec{k}_{||}$ and $\omega$. Thus, the stationary phase
method can be applied, and the dominant contribution to the coupling
comes from $\omega = E_F$ and $\vec{k}_{||}$ in the neighborhood of
points at which the argument of the exponential is stationary.  In
this limit both the sum in $\vec{k}_{||}$ and the energy integral in
Eq.\ref{j} can be evaluated analytically. The stationary points
$\vec{k}^0_{||}$ are the solutions of

\begin{equation}
\sum_\xi n_{\xi} \nabla k_\xi^\perp (\vec{k}_{||},E_{F}) = 0,
\label{j3}
\end{equation}
where $\nabla$ is the two-dimensional gradient in $k_{||}$ space.

For $\vec{k}^0_{||}$ belonging to sets 1 and 2 only one FS sheet
occurs in the first prismatic bulk BZ; the analysis then proceeds
exactly as in Ref. \onlinecite{VersL}. The corresponding periods are:
$p^b$=2.69 atomic planes and $p^n$=2.36 atomic planes respectively.
However, for the $M$-points the FS can be regarded as consisting of
two sheets. The two values of $k^\perp$ ($k_1^\perp$ and $k_2^\perp)$
associated with $\vec{k}^{0M}_{||}$, shown by arrows in Fig.
\ref{fig2}(b), correspond to equivalent periods
$p^n_1=2\pi/2k_1^\perp=10.97$ atomic planes and
$p^n_2=2\pi/2k_2^\perp=1.1$ atomic planes which cannot be
distinguished just by looking at discrete values of $N$.  However,
$\nabla k_1^\perp$ and $\nabla k_2^\perp$ vanish simultaneously when
calculated at $(\vec{k}^{0M}_{||},E_{F})$. Thus, Eq. \ref{j3} is
satisfied for any values of $n_1$ and $n_{2}$, yielding other periods
besides $p^n_1$ and $p^n_2$. The relative contributions of these
extrema depend on the comparative values of the corresponding
coefficients $c_{n_1,n_2}$.  The situation is very similar to that
discussed in Ref. \onlinecite{mauro}. Nevertheless, our calculations
have shown that the fundamental period $p^n_1$ (which is equivalent
to $p^n_2$) and its harmonics dominate. This is because the
coefficients associated with them are far larger than those
corresponding to alternative combinations of $n_1$ and $n_2$.

The calculated contributions to the coupling at T=300K coming
separately from each set of $\vec{k}^0_{||}$ are shown in Fig.
\ref{fig1}.  We note that all three contributions are comparable, but
the belly (full continuous line) clearly dominates. This contrasts
with FCC Co/Cu (001) trilayers where the belly contribution is
negligible in comparison with those given by the necks.\cite {PRL2}
The main reason for such difference is that minority carriers from
the vicinity of the FS belly are fully confined in BCC Cu by the Fe
layers in the ferromagnetic configuration. The physical origin of
such confinement is that the sp-like BCC Cu band which intersects the
FS has no counterpart at the minority-spin Fe FS. A similar situation
happens for the Cu FS states in the vicinity of $\vec{k}^{0}_{||}$
belonging to set 2, as shown in Fig.\ref{fig3}. On the other hand,
the Cu FS $M$-states of either spin can evolve into the corresponding
Fe FS states because they have sp-character, due to the existence of
a small but finite sp-d hybridization in this $k$-space region. The
degree of confinement experienced by the carriers in this case
depends on the relative sp-d hybridization strengths. The agreement
between the stationary phase approximation and numerical results
verifies that the exchange coupling at room temperature in Fe/Cu
(001) trilayers with perfect interfaces oscillates with a short
period strongly influenced by the belly contribution. This is in
accordance with the Philips group observations, as far as the period
is concerned. However, the calculated strength is an order of
magnitude larger than what they have observed. On the other hand the
amplitude of our long period component is about three times smaller
than the short period contribution.  We believe that the discrepancy
between experimental and theoretical results is due to interface
roughness which affects the amplitude and overall phase of the
coupling. \cite{inoue} The reason why the Philips results are much
smaller than those of the SFU group remains unexplained.

Motivated by recent measurements of the SFU group,\cite {CHC} we have
calculated the temperature variation of the coupling for different Cu
thicknesses.  Our results, shown in Fig.\ref{fig4}, are for perfect
interfaces where the bilinear exchange coupling $J_1$ is much larger
than the intrinsic biquadratic term. The rate of variation of
$|J_1(T)|$ with temperature clearly changes with spacer thickness.
The calculated slope for Fe/12Cu/Fe agrees very well with the
measured value for Fe/10Cu/Fe.  The most striking result however,
reproduced in the inset of Fig.\ref{fig4}, is the increase of the
coupling with increasing temperature for some Cu thicknesses. The
temperature dependence of $J_1$ is governed not only by the spacer FS
but also by the confining strength of the ferromagnetic layers
\cite{TD}, which differs for the three sets of extrema. As pointed
out in Ref.\onlinecite{TD}, the energy dependence of the phases
$\psi$ of the ``Fourier'' coefficients in Eq. \ref{j2} varies
according to the confinement strength and is very important in
determining the temperature dependence of the coupling. It turns out
that the values of $\partial \psi/\partial \omega$ calculated at the
second set of extrema are about four and a half times larger than at
the belly and the $M$-points. The temperature dependence of the
former contribution is then stronger than the others. Hence, the
coupling is approximately given by the sum of three oscillatory
functions of $N$ with comparable amplitudes which decay differently
with temperature.  Therefore, at some values of $N$, as in
Fe/13Cu/Fe, the balance is such that an overall increase in the
coupling is obtained even though the amplitude of each contribution
separately decreases with temperature.  Such increase was not
detected by the SFU group. One possible explanation is that they have
observed basically just a long period component. Another reason could
be the influence of spin fluctuations in the ferromagnetic layers,
which is neglected in our calculations.  Nevertheless, this is an
interesting temperature effect which may be observed under suitable
conditions.

\section{Acknowledgments}

We thank S.F. Pessoa for kindly providing us with the tight-binding
parameters of BCC Cu, D.M. Edwards for helpful discussions and T.J.P.
Penna for helping with the figures. This work has been financially
supported by CNPq and FINEP of Brazil, and SERC of U.K.

\begin{figure}
\caption{Calculated bilinear exchange coupling
at T=300K for BCC Fe/Cu/Fe (001) trilayer versus Cu thickness (solid
circles). The lines are contributions from the extremal points (see
text) corresponding to: the Cu FS belly (full line), the neck wave
vectors of set 2 (dashed line) and the neck M-points (dotted line).
The inset shows the total contribution from the three sets of 
extrema; the tick are the same as those of the main figure. }
\label{fig1}
\end{figure}

\begin{figure}
\caption{Calculated BCC Cu FS (a), and its relevant cross sections
for (001) (b). The arrows are the critical vectors
$k^{\perp}(\vec{k}^0_{||})$.}
\label{fig2}
\end{figure}

\begin{figure}
\caption{Band structures of bulk BCC Cu and Fe in the relevant
$[001]$ direction for the wave vectors $\vec{k}^0_{||}=(0,0)$ (a),
$\vec{k}^0_{||}=(0,0.324)$ (b) and $\vec{k}^0_{||}=(0.5,0.5)$ (c).}
\label{fig3}
\end{figure}

\begin{figure}
\caption{Calculated temperature dependence of the bilinear exchange
coupling for Fe/12Cu/Fe (solid circles) and Fe/14Cu/Fe (open
circles). The inset is for Fe/13Cu/Fe; the tick labels are the same
as those of the main figure. The lines are simply linear fits.}
\label{fig4}
\end{figure}

\end{document}